%% file: main.tex
\documentclass[journal]{IEEEtran}
\usepackage[utf8]{inputenc}
\usepackage{amsmath,amssymb,amsxtra,mathtools,bbm}
\usepackage{graphicx}
\usepackage{multirow,booktabs,makecell}
\usepackage{color,xcolor}
\usepackage{stfloats}
\usepackage{algorithm,algorithmic}
\usepackage{adjustbox}
\usepackage{tabularx,array,ragged2e}
\usepackage{subfig}
\usepackage{pgfplots}
\pgfplotsset{compat=newest}
\usepgfplotslibrary{groupplots}
\usepackage{xurl}
\usepackage{cite}
\usepackage[colorlinks=true,linkcolor=blue,citecolor=blue,urlcolor=blue]{hyperref}
\usepackage{comment}
\usepackage[utf8]{inputenc}
\usepackage{lscape} % put in preamble
\usepackage{longtable}

\graphicspath{{Figures/}}
\renewcommand{\arraystretch}{1.25}
\begin{document}

%\title{Parkinson's disease screening, diagnosis, prognosis and assessment of medical interventions via IoTs, wearables and contactless methods: state of the art, datasets and way forward in the gen AI era}

%\title{Towards Human-AI-Robot Collaboration for Parkinson's Disease Management: From Sensors to Closed-Loop Quality-of-Life Enhancement}
%\title{Towards AI-Agent based Closed-Loop Robotic Digital Twins for Parkinson’s Disease Management: A Review}
\title{Towards Human-AI-Robot Collaboration and AI-Agent based Digital Twins for Parkinson’s Disease Management: Review and Outlook}

\begin{comment}
Comprehensive and Scientific Tone "Wearables, Robots, and Intelligent Agents for Parkinson's Disease: Datasets, Technologies, and Closed-Loop Interactions" "A Systemic Review of IoT, Robotics, and AI Integration for Parkinson’s Disease Monitoring and Intervention" "Advancing Parkinson’s Disease Care through Sensor-Based Monitoring, Human–Robot Interaction, and AI Agents" "Parkinson's Disease in the Age of Intelligent Systems: A Review of Wearable Datasets, Robotics, and Multi-Agent Interactions" Human–Robot–AI Interaction Emphasis "Toward Closed-Loop AI Systems for Parkinson’s Disease: Wearable Sensing, Assistive Robotics, and LLM Integration" "Human–AI–Robot Collaboration for Parkinson’s Disease: From Sensors to Closed-Loop Quality-of-Life Enhancement" "Wearables to Robots: A Unified Framework for AI-Assisted Parkinson's Disease Monitoring and Care" Dataset-Driven Emphasis "A Survey of Wearable Sensor Datasets and Multi-Agent Interventions for Parkinson's Disease" "Data-Driven Parkinson’s Disease Care: Review of Wearable Sensing, LLM Agents, and Human–Robot Feedback Loops" AR/VR and Future-Oriented Tone "Immersive and Intelligent Technologies for Parkinson’s Disease: Toward AR/VR, LLMs, and Multi-Agent Feedback Ecosystems" "Future Directions in Parkinson’s Disease Management: A Cross-Modal Survey of Sensors, Robots, LLMs, and Human–AI Loops"

\end{comment}

\author{
%\IEEEauthorblockN{
Hassan Hizeh,~\IEEEmembership{Member,~IEEE}, Rim Chighri,~\IEEEmembership{Member,~IEEE}, Muhammad Mahboob Ur Rahman,~\IEEEmembership{Senior Member,~IEEE}, Mohamed A. Bahloul,~\IEEEmembership{Senior Member,~IEEE}, Ali Muqaibel,~\IEEEmembership{Senior Member,~IEEE}, and Tareq Y. Al-Naffouri,~\IEEEmembership{Fellow,~IEEE}
\thanks{ 
Hassan Hizeh, Muhammad Mahboob Ur Rahman and Tareq Y. Al-Naffouri are affiliated with Computer, Electrical and Mathematical Sciences and Engineering Division (CEMSE), King Abdullah University of Science and Technology, Thuwal 23955, Saudi Arabia. 
Emails: \{hassan.hizeh,muhammad.rahman, tareq.alnaffouri\}@kaust.edu.sa. \\
Mohamed A. Bahloul and Rim Chighiri are affiliated with the College of Engineering \& Advanced Computing and the Translational Biomedical Engineering Research lab, Alfaisal University, Riyadh, 11533, Saudi Arabia. Email: \{mbahloul, rchighri\}@alfaisal.edu.\\
Ali Muqaibel is affiliated with the Electrical Engineering Department and the Center for Communication Systems and Sensing, King Fahd University of Petroleum and Minerals (KFUPM), Dhahran 31261, Saudi Arabia. Email: muqaibel@kfupm.edu.sa. 
\\
The research reported in this publication was supported by funding from King Abdullah University of Science and Technology (KAUST) - KAUST Center of Excellence for Smart Health (KCSH), under award number 5932. 
} 
}

\vspace{-0.8cm}

\maketitle

\begin{abstract}

The current body of research on Parkinson's disease (PD) screening, monitoring, and management has evolved along two largely independent trajectories. The first research community focuses on multimodal sensing of PD-related biomarkers using noninvasive technologies such as inertial measurement units (IMUs), force/pressure insoles, electromyography (EMG), electroencephalography (EEG), speech and acoustic analysis, and RGB/RGB-D motion capture systems. These studies emphasize data acquisition, feature extraction, and machine learning-based classification for PD screening, diagnosis, and disease progression modeling. In parallel, a second research community has concentrated on robotic intervention and rehabilitation, employing socially assistive robots (SARs), robot-assisted rehabilitation (RAR) systems, and virtual reality (VR)-integrated robotic platforms for improving motor and cognitive function, enhancing social engagement, and supporting caregivers.
Despite the complementary goals of these two domains, their methodological and technological integration remains limited, 
%The sensing community typically stops at detection or assessment, while the robotics community focuses on actuation and assistance, 
with minimal data-level or decision-level coupling between the two. With the advent of advanced artificial intelligence (AI), including large language models (LLMs), agentic AI systems, a unique opportunity now exists to unify these disconnected research streams.
We envision a closed-loop sensor-AI-robot framework in which multimodal sensing continuously guides the interaction between the patient, caregiver, humanoid robot (and physician) through AI agents that are powered by a multitude of AI models such as robotic and wearables foundation models, LLM-based reasoning, reinforcement learning, and continual learning. Such closed-loop system enables personalized, explainable, and context-aware intervention, forming the basis for digital twin of the PD patient that can adapt over time to deliver intelligent, patient-centered PD care. 

\end{abstract}
%\vspace{-0.6cm}

\section{Introduction}
Parkinson's disease (PD) is a progressive neuro-degenerative disease that is characterized by motor symptoms such as tremor, rigidity, bradykinesia, paucity, akinesia, absence of movement, freezing of gait and postural instability \cite{sigcha2023deep}. A patient's quality of life is severely impacted by these motor symptoms, which get worse over time \cite{zhao2021quality}. While PD is not a direct cause of death, it leads to a higher risk of mortality in patients by putting them at the risk of nervous, circulatory, respiratory, and infectious diseases \cite{ryu2023mortality}. PD poses a great burden to the socioeconomic status of a country with PD patients doubling over the past generation \cite{dorsey2018global}. The wide array of both motor and non-motor symptoms of PD pose a substantial and pressing management challenge for medical professionals and engineers alike.

For a patient to be diagnosed with PD, they must exhibit bradykinesia in addition to either rigidity or tremor \cite{kobylecki2020update}. The movement disorder society-sponsored revision of the unified Parkinson's disease rating scale (UPDRS) is the internationally accepted clinical gold standard for assessing PD severity and progression. It is a comprehensive tool comprising four parts: i) non-motor experiences of daily living, ii) motor experiences of daily living, iii) motor examination, and iv) motor complications \cite{goetz2008movement}. Each item is rated on a 0-4 scale of increasing severity. Administered by trained clinicians, the UPDRS provides a structured quantitative framework for evaluating the multifaceted motor and non-motor symptoms of PD, enabling the standardized tracking of disease course and therapeutic efficacy across clinical and research settings. Another rating system is the Hoehn \& Yahr system that is used to describe the progression of motor symptoms in Parkinson's disease. It ranges from stage 0 (no signs) to stage 5 (wheelchair-bound or bedridden unless aided). However, it is important to note that both the UPDRS and Hoehn \& Yahr system are rating scales, and not a diagnostic tool.
%UPDRS assesses motor experiences of daily living, non-motor experiences of daily living, motor examination, and motor complications, is most frequently used in clinical assessment \cite{goetz2008movement}. 

Men are about 1.4 times more likely than women to have PD \cite{dorsey2018global}. Molecular biology research has shown that PD is marked by the buildup of $\alpha$-synuclein protein (forming Lewy bodies and Lewy neurites) and the loss of dopamine-producing neurons in a brain region called the substantia nigra pars compacta \cite{kouli2018parkinson}. Therefore, neuropathological confirmation is necessary for a definitive diagnosis; however, brain biopsy is not feasible in routine practice due to its high invasiveness and significant risks \cite{warren2005brain}. %Therefore, while we wait for proven disease-modifying treatments, current management places a higher priority on function and quality of life.

A definitive clinical diagnosis of Parkinson's disease often occurs very late, potentially 12 to 14 years after the disease process has begun \cite{postuma2012identifying}. Current evidence suggests that the disease may start in the peripheral autonomic nervous system and progress to the lower brainstem structures before affecting the substantia nigra pars compacta \cite{kouli2018parkinson,chen2020autonomic}. This established pathology explains the presence of subtle prodromal symptoms, such as depression, tremor, fatigue, constipation, dizziness, and balance problems, that are observed years before the official diagnosis is made \cite{schrag2015prediagnostic}.

There is currently no established biomarker or conclusive test for early diagnosis of Parkinson's disease, despite its clinical utility. 
The primary method of (late) diagnosis is clinical examination, which is subject to classification errors by neurologists. 
Clinical tests like the timed up-and-go (TUG) test are frequently used to identify motor impairments early on. The TUG test measures the time it takes for a patient to stand from a seated position, walk three meters, turn, walk three meters again, and sit back down \cite{williams2018association}. However, tests like TUG, record only a single moment in time, and thus, are episodic in nature.

%This tension between complex, fluctuating symptoms and episodic assessment drives researchers to develop new methods that can sense, interpret, and respond between clinic visits. 
%That is why, new sensing and artificial intelligence (AI) techniques seek to bring that perspective into daily life.

The current PD management techniques place a higher priority on symptomatic relief, function and quality-of-life of PD patient. Levodopa and other dopamine replacement treatments serve as the cornerstone of treatment, while other pharmaceuticals are recommended to treat non-motor symptoms like dementia and depression \cite{kouli2018parkinson}. However, since patients are usually evaluated every six to nine months, it is difficult to identify minute, transient changes in the course of the disease. Thus, PD care still depends on quick clinic snapshots rather than ongoing, in-person observation, which is clearly a sub-optimal treatment methodology.

%An additional justification for richer, continuous data is that these symptomatic strategies are most effective when timing and dosage can be adjusted to a patient's daily variability.

\begin{figure*}
    \centering
    \includegraphics[width=1\linewidth]{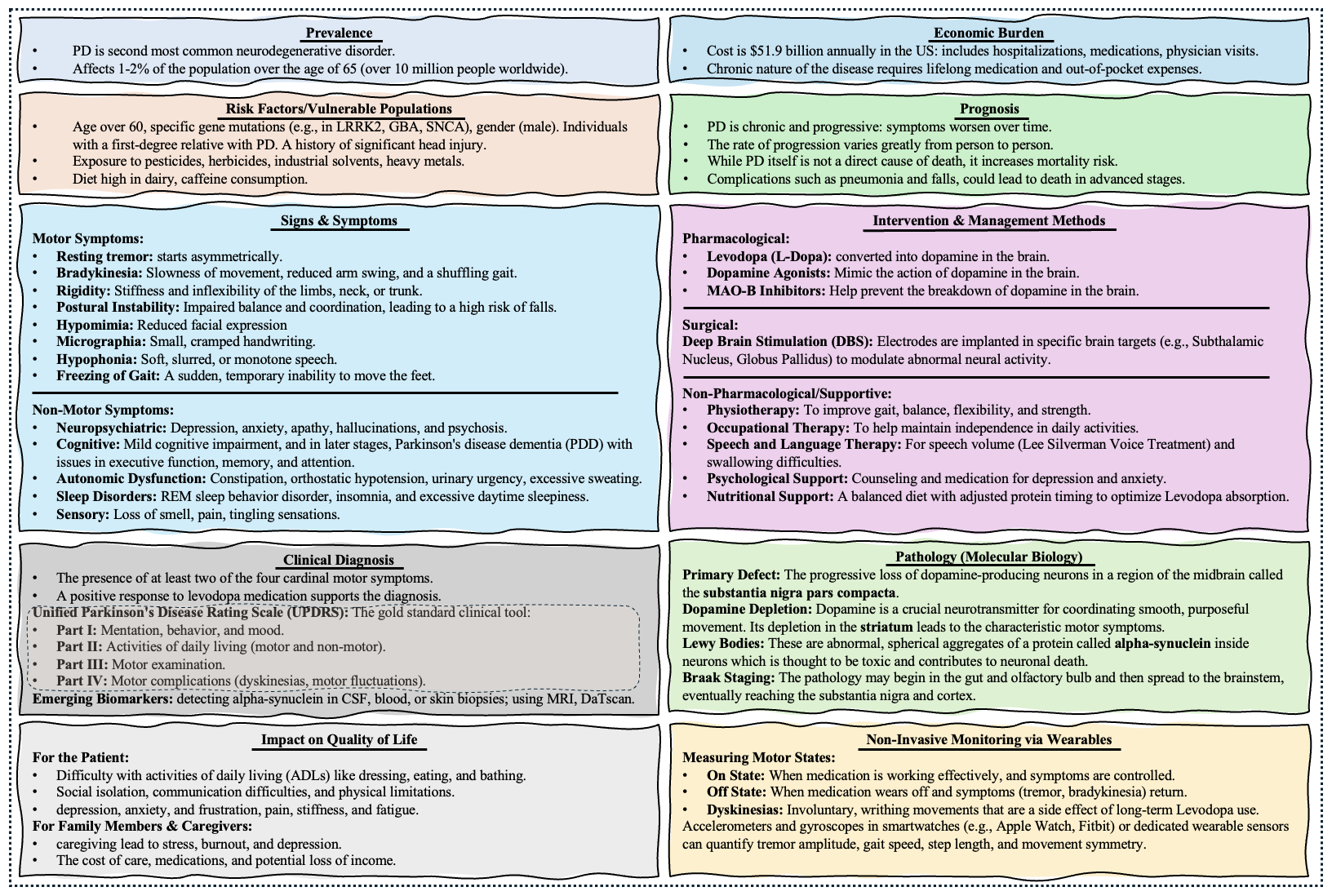}
    \caption{PD in a nutshell: prevalence, economic burden, vulnerable populations, risk factors, signs and symptoms (motor and non-motor), impact on quality of life of patient and that of family members and caregivers, pathology (based on molecular biology), biomarkers and clinical diagnosis (UPDRS and more), prognosis, intervention methods, non-invasive monitoring via wearables for measuring on/off states for measurement of effectiveness of medical intervention.}
    \label{fig:pd1}
\end{figure*}

Fig. \ref{fig:pd1} provides a compact summary of various aspects and traits of PD, i.e., its prevalence, economic burden, vulnerable populations, risk factors, signs and symptoms (motor and non-motor), impact on quality-of-life of patient and that of family members and caregivers, pathology, biomarkers and clinical diagnosis, UPDRS rating system, prognosis, intervention methods, non-invasive monitoring via wearables for measuring on/off states for measurement of effectiveness of medical intervention.

\textbf{Scope of this survey paper:}
Research efforts for PD diagnosis, monitoring, intervention, and prognosis are multidisciplinary and fragmented, 
with distinct research communities advancing parallel yet largely disconnected lines of inquiry. Broadly speaking, four major domains dominate the current PD research ecosystem. The first comprises researchers focused on sensing technologies, who have developed and validated a variety of noninvasive modalities, such as inertial measurement units (IMUs), force or pressure insoles, electromyography (EMG), electroencephalography (EEG), and speech or acoustic analysis, to enable continuous monitoring of motor and non-motor symptoms, moving beyond the traditional reliance on episodic clinical assessments \cite{pasluosta2015emerging}. A second research community centers on robotic systems, particularly socially assistive robots (SARs) and robot-assisted rehabilitation (RAR) platforms, which aim to provide motor rehabilitation, cognitive support, and social engagement for PD patients \cite{perju2022artificial}. In parallel, molecular biologists and neuroscientists focus on elucidating the underlying pathological mechanisms of PD, including alpha-synuclein aggregation, mitochondrial dysfunction, and neuroinflammatory processes, with the goal of identifying reliable biomarkers and therapeutic targets \cite{simon2020parkinson}. Complementing these efforts, pharmacological researchers strive to design and optimize novel medications and drug delivery systems to slow disease progression and improve symptom management \cite{connolly2014pharmacological}. While each of these communities contributes critical insights to the broader understanding and management of PD, methodological and technological silos persist, limiting cross-domain integration. 

This survey paper aims to review the literature from the first two research communities, i.e., sensing and robotics. We believe it is high time for researchers to begin efforts to bridge these two research streams, particularly through advances in artificial intelligence (AI), data fusion, and closed-loop modeling. That is, there exists a unique opportunity to create unified, intelligent systems capable of linking sensing, AI and robotic intervention for patient-centered PD care.

\textbf{Limitations of existing survey papers:}
There do exist a number of survey papers on Parkinson's disease screening, monitoring, management and prognosis. However, a key limitation of the previous survey papers is that each such paper originates either from the research community of sensors \cite{rovini2017wearable} or from that of robotics \cite{tao2024role}. Therefore, each of the previous survey papers provides a localized view of the PD research landscape. More precisely speaking, the first set of survey papers focus on development of sensors, wearables, internet of things (IoT) and internet of medical things (IoMT) modules, and data acquisition from PD patients and healthy controls \cite{mughal2022parkinson}. On the other hand, the second set of survey papers focus on how various kinds of robots, e.g., SARs could help improve the quality-of-life of PD patients \cite{bar2023socially}. 

This survey paper intends to unify the disparate discussions from the sensors community and robotic community, and to provide a vision for the future PD research, in the era of emerging concepts such augmented reality/virtual reality (AR/VR), emerging AI tools such as large language models (LLMs), foundation models for robots and sensors, reinforcement learning, continual learning, meta learning, agentic AI, AI agents, quantum AI, and physical AI. 
%Together, these could help realize a PD digital twin. 

Table \ref{tab:survey_comparison} compares this survey paper with previous reviews in three aspects: scope, modality coverage, and whether or not the discussion in the previous reviews covers the literature from both sensors and robotics communities along with their potential integration into an AI-powered closed-loop system. 

\begin{table*}[t]
\centering
\caption{Comparison of this survey paper with some recent survey papers on PD}
\label{tab:survey_comparison}
\renewcommand{\arraystretch}{1.1}
\setlength{\tabcolsep}{4pt}
\begin{tabularx}{\textwidth}{p{1.5cm}X X X}
\toprule
\textbf{Ref.} & \textbf{Scope} & \textbf{Modalities} & \textbf{Integration / Closed-loop} \\
\midrule

\cite{sigcha2023deep} & PD diagnosis, monitoring, prognosis (motor + non-motor) & inertial, force, mic, EEG/EMG & No \\
\cite{brognara2019assessing} & Gait assessment & IMU only & No \\
\cite{rovini2017wearable} & Diagnosis \& monitoring & IMU, EEG, EMG & No \\
\cite{sapienza2024assessing} & Monitoring, clinical utility, user experience & IMU-based wearables in daily life & No \\
\cite{rabie2025review} & Detection \& progression & ML/DL methods & No \\
\cite{balakrishnan2022role} & Diagnosis/monitoring via gait & IMU, insole force/pressure & No \\
\cite{sica2021continuous} & Free-living gait monitoring & IMUs & No \\
\cite{zhang2024detection} & FoG detection \& prediction & IMU, smartphone, EEG/EMG & No \\
\cite{di2024machine} & PD vs healthy discrimination & IMU, smartphone & Discusses future integration with EHR \\
\cite{moreau2023overview} & Real-world monitoring \& therapy adjustment & Wearable IMUs & No \\
\cite{perju2022artificial} & Home-based therapy mgmt, telemedicine, robot-assisted training & Robotic gait systems, CV+ML & LCIG titration by telemedicine (remote dose adjustment) \\
\cite{bougea2025application} & Real-time monitoring (gait, tremor, dyskinesia) & IMUs, speech, vision, EEG/EMG, biosensors, smartphones, smart-home, e-skin & No \\
\textbf{This work} & Diagnosis, monitoring, rehab, assistive & IMU, multimodal sensors, SAR/RAR robots, LLMs & Yes \\
\bottomrule
\end{tabularx}
\end{table*}

\textbf{Contributions:}
Compared to existing survey papers, this survey paper bridges the following gaps in the literature:
\begin{itemize}
    \item This paper covers a very wide spectrum of PD research, from risk screening and diagnosis through longitudinal monitoring and therapy support. However, most of the existing surveys focus on just a few aspects of PD only, e.g., screening, monitoring, early detection, or prognosis.
    \item This work is the first to synthesize the literature across three distinct and isolated research communities: 1) sensors such as wearables (including IMUs and other biosignals), speech and vision sensors, 2) robotic systems, and 3) emerging AI tools (LLMs and agentic frameworks). Previous work usually focuses on one research domain at a time (wearables/IMUs, or rehabilitation robotics, and very rarely LLMs/AI agents). 
    \item Previous reviews treat sensing and robotic intervention as distinct elements. A major motivation of this work is to highlight the emerging potential for a closed-loop architecture that connects multimodal sensing, AI-based inference, and adaptive robotic intervention. We foresee a closed-loop system whereby sensors continuously monitor the behavior and physiology of the PD patient, AI models (agentic AI, LLMs, reinforcement learning) contextualize the sensed data and update the PD digital twin, and a therapeutic module (such as a robot) adjusts the intervention in real time, then learns from the results to improve the subsequent action. In PD, where symptoms can vary (e.g., ON/OFF medication states, episodic freezing-of-gait), this paradigm works well because prompt feedback can enhance motor function or lessen the severity of symptoms.
\end{itemize}

%\textbf{Contributions:}
%This paper makes the connection between technological potential and clinical need. We review the machine-learning techniques developed on top of the sensing modalities commonly used in PD, including wearable IMUs, EEG/EMG, speech/voice, and video, and we look at how robotic systems convert algorithmic insight into tangible therapy. 
%We also present new AI frameworks that can close the loop between measurement and intervention, personalize decisions, and fuse heterogeneous data, such as digital twins, agentic AI, and LLMs. Our main focus is on the importance of a closed-loop strategy. 

\textbf{Literature Review Criterion:} 
To search for the relevant literature, we scoured various academic research databases, i.e., Scopus, Web of Science, PubMed, IEEE Xplore, ScienceDirect, Google Scholar, Arxiv, and other online resources. Specifically, we searched the relevant literature spanning the years from 2003 to 2024. To retrieve the published articles of interest, we utilized the following keywords in various combinations: Parkinson’s disease; wearables; IMU; EEG; EMG; speech; computer vision; rehabilitation robotics; socially assistive robots; virtual reality; digital twin; large language models; AI agents; closed-loop care. We focused particularly on highly cited and influential works. 
%After gathering the relevant literature, we categorized the works based on the sensing modality used.

\begin{figure}
    \centering
    \includegraphics[width=1\linewidth]{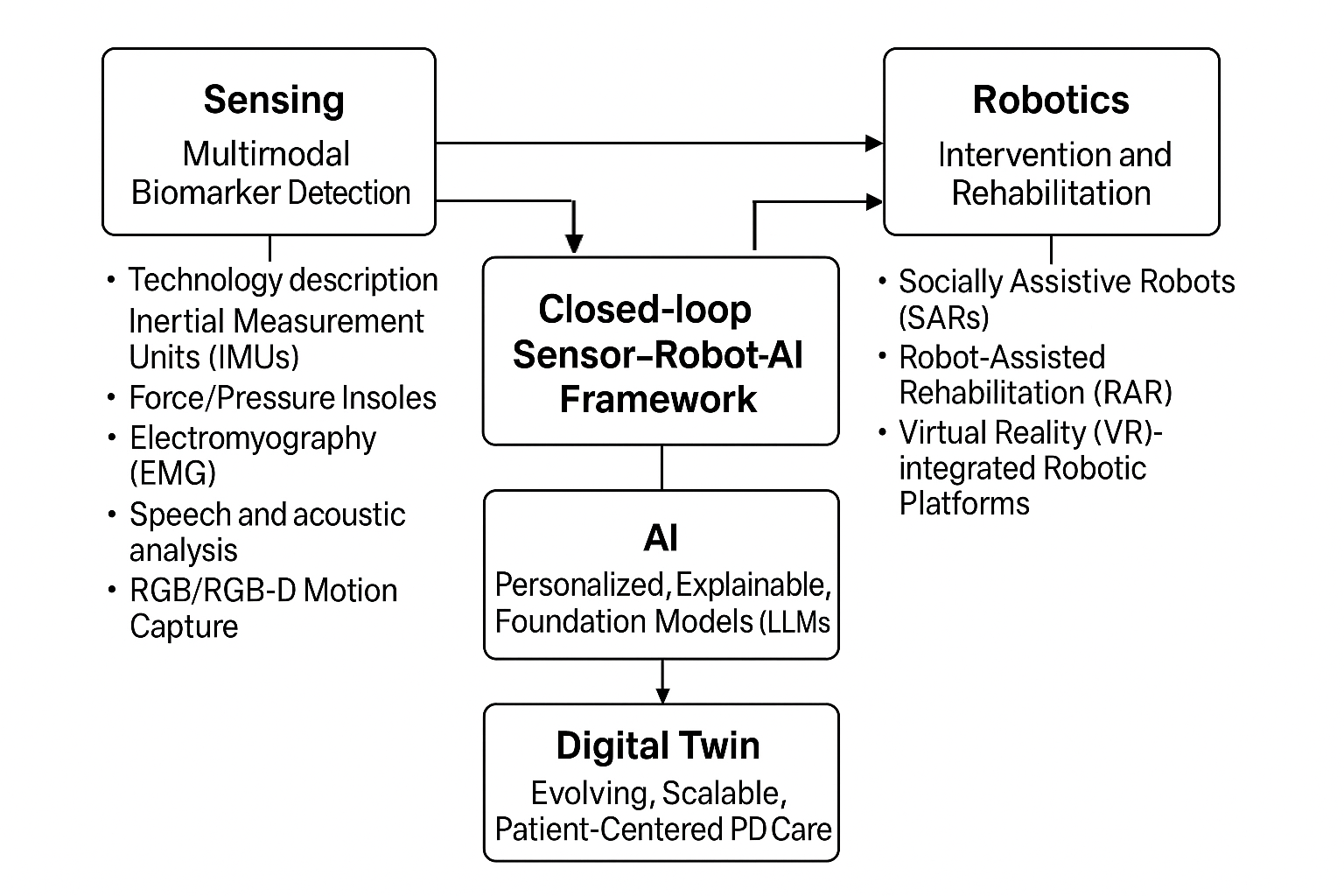}
    \caption{Outline of this survey paper: Section II reviews selected literature that utilizes various sensors for PD research. Section III discusses the state-of-the-art on robots-based PD research. Section IV highlights the potential of closed-loop integration of sensor and robots via advanced AI tools such AI agents, LLMs, reinforcement learning to realize the concept of PD digital twin.}
    \label{fig:outline}
\end{figure}

\textbf{Paper Outline:}
Fig. \ref{fig:outline} provides a graphical outline of this survey paper. That is, Section II reviews the literature from the sensors community that forms the first research pillar of PD research. Section III reviews the literature from the robotics community that makes the second research pillar of PD research. Section IV discusses the role of emerging AI frameworks (LLMs, AI agents, reinforcement learning, robotic foundation models) in PD research, and articulates the integration of sensors-AI-robots in a closed-loop to establish a PD digital twin, as envisioned by the authors. 

%\textbf{Outline:}
%Fig. \ref{fig:outline} provides a graphical outline of this survey paper. That is, Section II reviews selected literature that utilizes various sensors for PD research. Section III discusses the state-of-the-art on robots-based PD research. Section IV highlights the potential of closed-loop integration of sensor and robots via advanced AI tools such AI agents, LLMs, reinforcement learning to realize the concept of PD digital twin.

\section{Research Pillar I: Wearable and IoT Sensors for Parkinson's Disease Care}

As mentioned earlier, this section summarizes the first set of research literature that focuses on development of sensors, wearables, IoT and IoMT modules and data acquisition from PD patients and healthy controls under a wide range of settings. Such works also come under the umbrella of human activity recognition (HAR) problem, when applied to PD patients. This section summarizes selected related work for each sensing modalities, as well as the works that utilize multiple sensors to collect multi-modal data. Finally, we also provide a detailed discussion of the prevailing situation from the perspective of AI, i.e., public and private datasets available for PD research, their sample sizes, labeling methodology, training and validation methods, and more. 

\subsection{Motion Cameras}

Motion sensors based on cameras have been extensively studied as non-invasive and affordable methods of PD screening. The Microsoft Kinect V2, a depth-sensing camera, was used in early research to examine how PD patients' and healthy controls' gaits differed during the TUG test. Although the system was able to differentiate between patients and controls, its small sample size (19 patients and 8 controls) limited generalizability and its low sensitivity made it impossible to detect minute changes associated with medication \cite{van2021camera}. The potential of motion analysis for objective quantification was highlighted by a later Kinect study that evaluated bradykinesia across five standardized tasks and discovered strong correlations with clinical rating scales \cite{wu2023kinect}.

The focus of other studies has shifted to more widely available hardware. Standard webcams have demonstrated good agreement with wearable IMUs and can track upper-limb bradykinesia with high accuracy when combined with AI algorithms. These inexpensive configurations hold great promise for decentralized trials and telemedicine \cite{monje2021remote}. More sophisticated applications, like VisionMD, showed that software-driven video analysis can detect drug effects more accurately than depth cameras, indicating that algorithmic sophistication might be more significant than specialized hardware \cite{lange2025computer}.

Multi-camera systems like TULIP use six synchronized cameras to create 3D reconstructions and extract minute disease-specific signatures in order to further increase accuracy. Despite their great efficacy, these systems are complicated and unfeasible outside of research settings \cite{kim2024tulip}. Although hybrid setups, like the ReadyGo system that combines RGB and depth sensors, can predict motor severity scores and distinguish early-stage PD from healthy controls with 91\% accuracy, their comprehensiveness is limited by their reliance on gait parameters alone \cite{yin2024gait}.

Additionally, machine learning frameworks have been created to directly extract features from unprocessed clinical videos. This allows for automated UPDRS score prediction and even the discovery of new disease markers in the movements of the fingers and lower limbs \cite{deng2024interpretable,dooley_2025}. Lastly, vision-based sensors and cloud computing are combined in embedded real-time systems like VPDIS to stage the severity of PD and give clinicians feedback. Despite their effectiveness, these methods raise issues with energy consumption, privacy, and reliance on network connectivity \cite{jinila2022vision}.

In short, camera-based systems provide a flexible, contactless, and scalable way to record PD motor symptoms. Although there are still issues with sensitivity, accessibility, and data privacy, they have a lot of promise for unsupervised home-based monitoring, when paired with wearable sensors and cutting-edge AI techniques that improve accuracy and lessen clinical subjectivity \cite{vun2024vision}.

% ====== Where you want the multi-page table ======
%\clearpage
%\onecolumn
%\newgeometry{left=1.5cm,right=1.5cm,top=1.8cm,bottom=1.8cm} % adjust if you like
%\setlength{\LTleft}{1pt}
%\setlength{\LTright}{1pt}
%\setlength{\LTcapwidth}{\textwidth} % caption width = table width

% IMPORTANT: do NOT wrap longtable in table/table*, center, minipage, etc.
%\footnotesize
%\begin{longtable}{@{} P{0.18\textwidth} P{0.26\textwidth} P{0.26\textwidth} P{0.26\textwidth} @{}}

%\restoregeometry
%\twocolumn

\subsection{Voice and Speech Analysis Sensors}

One useful method for examining how PD affects speech, one of the first motor symptoms that may manifest years before a diagnosis is made, is the use of acoustic sensors \cite{rusz2021speech}. These sensors have the potential to be used for early screening and disease monitoring because they can identify subtle vocal impairments, even though they are unable to detect classic motor signs like bradykinesia, rigidity, or tremor.

Three types of vocal characteristics are usually the focus of this research: prosodic (e.g., pitch, maximum phonation time, speaking rate), articulatory (e.g., Mel-Frequency Cepstral Coefficients [MFCC], Cepstral Peak Prominence [CPP], vowel articulation index, Perceptual Linear Prediction Coefficients [PLPC]), and phonatory (e.g., jitter, shimmer, Harmonics-to-Noise Ratio) \cite{wright2025vocal}. The most extensively researched of these is sustained vowel phonation. For example, Tsanas et al. recorded 42 patients' prolonged "ahh" sounds using a home-based telemonitoring system equipped with headset microphones. Their analysis showed a strong correlation between dysphonia and the progression of Parkinson's disease \cite{tsanas2009accurate} and showed that speech features could estimate UPDRS ratings. Even though they work well, these configurations are still expensive, fixed, and reliant on operator skill, which restricts scalability.

Researchers have utilized the microphones built into smartphones and tablets and have demonstrated that even consumer-grade hardware can approximate the performance of laboratory equipment with appropriate noise control and algorithmic adaptation. The viability of mobile telemedicine platforms was highlighted by Illner et al. and Rusz et al., who showed that smartphone-based articulation and tone analysis could attain near-clinical accuracy when compared with professional condenser microphones \cite{rusz2018smartphone,illner2024smartphone}. Although these methods improve patient comfort and accessibility, they are limited by hardware variability, which could make it more difficult to identify subtle impairments.

In summary, speech sensors can detect common speech abnormalities associated with PD, including monotonicity, reduced loudness, and dysarthria. They are useful for tracking the course of a disease and for pre-diagnostic monitoring. Voice-based methods are best viewed as supplementary rather than independent diagnostic tools, though, as not all patients show speech abnormalities and speech analysis does not encompass the entire range of PD motor symptoms \cite{wright2025vocal}.

\subsection{Electrophysiological Sensors}

As an objective, quantitative biomarkers for PD, electrophysiological sensors such as electroencephalography (EEG), electromyography (EMG), and electrocardiography (ECG) have long been investigated. Of these, EEG and EMG are most frequently used in wearable technology and provide information on both motor abnormalities and cortical dysfunction.

\subsubsection{Electroencephalography (EEG) Sensors}

EEG sensors record electrical activity in the brain and are frequently used to investigate cortical alterations linked to PD. EEG-derived somatosensory evoked potentials (SEPs) offer indicators of motor activity phases, and elevated beta-band power is a consistent characteristic of advanced PD \cite{molcho2023evaluation}. Molcho et al. showed that cognitive EEG features that correlated with F-DOPA PET imaging could be used to classify PD patients using low-channel wearable EEG and machine learning, indicating the possibility of portable systems centered on cognitive biomarkers \cite{molcho2023evaluation}. Low-channel devices, however, have poor spatial resolution. More accuracy can be obtained with high-density EEG with sophisticated signal processing. A 2022 study that used machine learning and common spatial pattern filtering was able to distinguish off-medication PD patients from controls with 99\% accuracy, with alpha- and beta-band features showing the highest discriminative power \cite{aljalal2022parkinson}. Motion artifacts and poor signal quality outside of clinical settings continue to be significant obstacles in spite of these advancements. Emotional EEG paradigms have been tested more recently, showing near-perfect classification between PD and healthy subjects and suggesting non-motor biomarkers that are ecologically valid \cite{pasmanasari2021potential}.

\subsubsection{Electromyography (EMG) Sensors }

EMG sensors measure muscle activity directly and are particularly good at identifying the motor symptoms of PD, including tremor, stiffness, and irregular gait. Adhesive electrodes are commonly used on muscles like the tibialis anterior, gastrocnemius, and soleus in surface electromyography (sEMG). For instance, using standard signal processing, Putri et al. were able to distinguish PD patients from controls with 97\% accuracy by recording voluntary contractions during gait \cite{putri2016electromyography}. In lab settings, EMG is a dependable biomarker due to its strong signal quality and high reproducibility. However, its applicability for unsupervised or home-based screening is limited by practical issues such as electrode placement and need for calibration.

Multi-channel synergy analysis, made possible by advancements in electrode arrays, has shown that PD patients have altered muscle coordination, with partial normalization occurring under pharmacological therapy \cite{lanzani2025methodological}. These configurations are less portable, and wireless transmitters are integrated into such systems for long-term data collection during daily activities \cite{lu2020evaluation}. However, motion artifacts, irregular electrode-skin contact, battery limitations, and signal deterioration outside of controlled environments present some significant limitations. Nevertheless, such systems have the potential to greatly enhance early detection and progression monitoring of PD.

\subsection{Pressure/Force Sensors}

In PD screening, force and pressure sensors, which are usually found in instrumented insoles, mats, or skin-mounted arrays, are frequently used, especially to measure tremor and abnormal gait. These systems record spatiotemporal gait parameters that are used in conjunction with clinical evaluations, and their potential for monitoring and diagnosis is being assessed more and more.
In addition to contact-based insoles and mats, ambient structural vibration sensing can infer joint kinematics without the need for body-worn devices; in a study involving 20 participants, Dong et al. model human–structure interactions using a graph transformer informed by physics and report 3.9° MAE over 12 joint angles, suggesting a path for home gait analysis that preserves privacy \cite{Dong2025PIGT}. 

In a large study, a dual-stage machine learning model was used to analyze vertical ground reaction force (VGRF) signals from both feet in 166 participants (93 PD, 73 controls). The accuracy of the system was 96.4\% for severity assessment and 97.5\% for PD detection \cite{navita2025gait}. High sensitivity was achieved through sophisticated feature selection and preprocessing techniques, but generalizability to real-world situations is limited by reliance on treadmill-based laboratory testing.

For longitudinal monitoring, portable smart insoles have been investigated in smaller studies. Pressure-derived gait metrics demonstrated reliable differentiation between ON and OFF medication states and strong correlations with UPDRS-III scores in a group of 14 PD patients \cite{Karamanis2024}. Repeatability across sessions was shown by these insoles, indicating that they may have use in monitoring medication response. Subgroup analyses were less robust, though, due to the small sample size and underrepresentation of patients in advanced stages.

Similarly, in 25 PD patients, Parati et al. verified pressure-sensitive insoles against a gold-standard clinical walkway in both single- and dual-task scenarios. The insoles' wireless integration enhanced user convenience, and they demonstrated moderate-to-excellent agreement with the reference system \cite{parati2022}. However, external validity is once more constrained by the lack of healthy controls and the clinic-only testing setting.

All things considered, pressure and force sensors offer repeatable, objective measurements for PD gait analysis that have been shown to correlate with positive clinical outcomes. Although promising for long-term and home-based monitoring, widespread adoption requires validation in free-living settings and larger cohorts.

\subsection{Radio Signal Sensors}
With benefits over conventional wearable technology, radio signal sensors have recently become a cutting-edge, non-contact method for Parkinson's disease screening and monitoring. These gadgets, which are frequently about the size of a Wi-Fi router, send out low-power radio waves into the surrounding air and then use the signals that are reflected from the body to determine physiological indicators like breathing patterns, movement speed, and gait dynamics while you sleep. They are ideal for long-term, unsupervised use because they don't require physical contact, device management, or behavior modification like accelerometers or EMG do.

In a groundbreaking study, Yang et al. used radio wave sensors to assess the nocturnal breathing patterns of over 7,600 people. The system outperformed wearable breathing belts used for single-night monitoring, detecting PD with an area under the curve (AUC) of 0.91 and increasing sensitivity to 95\% when multiple nights of data were analyzed \cite{Yang2022AIbreathing}. This strategy's main advantages were its high adherence, passive operation, and adaptability to domestic settings. Among the drawbacks were the requirement for strong signal processing to separate signals from the target person and the possibility of interference from outside elements like fans or pets.

In a year-long study, Liu et al. collected over 200,000 gait speed measurements from 34 PD patients and 16 controls, further showcasing the potential of these sensors \cite{liu_zhang2022}. The system allowed for extremely sensitive, ongoing monitoring of the course of the disease, identifying minute variations and drug reactions that are frequently overlooked in evaluations conducted in a clinic. Since there was no need for interaction with the device, compliance was almost perfect; however, multi-occupant households presented difficulties, requiring algorithms to accurately attribute signals to the intended patient.

All things considered, radio signal sensors offer a safe and contactless way to keep an eye on PD. They stand out from conventional wearables due to their high patient adherence, ecological validity, and ease of use. To facilitate integration into standard clinical practice, future research should concentrate on improving spatial resolution, handling multi-occupant environments, and developing miniaturization.

\subsection{Inertial measurement units (IMUs)}

Recent advancements in technology have produced small, reasonably priced, and discrete sensing devices that have great potential to revolutionize clinical rehabilitation procedures\cite{porciuncula2018wearable}. IMUs are one type of sensing technology that is becoming more and more significant in the assessment of movement disorders and rehabilitation.
Compact, wearable electronic devices called IMUs that use integrated triaxial accelerometers, gyroscopes, and magnetometers to measure linear acceleration, angular velocity, and frequently magnetic field orientation. These sensors are crucial for real-time estimation of body posture, movement, and orientation because they allow motion reconstruction in three dimensions \cite{sy2022engineer}.
 
Wearable IMUs allow for continuous, objective monitoring of PD motor function, capturing subtle fluctuations and variations that occur throughout daily life and providing data in near real time, outside of the clinic, in contrast to traditional clinical assessments like UPDRS, which only provide intermittent and subjective snapshots of motor performance\cite{ossig2016wearable, sica2021continuous}. To elaborate, waist accelerometry has the potential to be a low-cost screening tool and to enrich trials because, at the population level, it can identify prodromal PD up to about seven years before a clinical diagnosis and beats models based on genetics, way of life, blood biochemistry, or prodromal symptom questionnaires \cite{Schalkamp2023NatMed}.
 
Gait abnormalities and kinematic features, such as gait cycle duration, step length, and walking speed are commonly analyzed to differentiate individuals with PD from healthy controls\cite{sica2021continuous}. Also, they are extensively used in IMU- based Parkinson disease research to objectively detect other symptoms, including tremor, bradykinesia, dyskinesia, and mobility-related impairments, providing reliable and continuous clinical measurement\cite{sica2021continuous,dai2015tremor,dai2015bradykinesia}.
 
IMUs can be worn or securely attached to various parts of the body, including the feet, wrists, waist, chest, and head, allowing for the collection of detailed movement data from multiple regions. Their flexible placement enables the extraction of a wide range of biomechanical and kinematic parameters such as step length, gait speed, joint angles, shoulder and limb rotation, and postural alignment.

From the IMU produced raw time-series signals, pertinent motor signals are obtained through filtering. Then, after segmentation, and windowing, features are typically extracted in the time, frequency, and spatiotemporal domains. For instance, band-pass filters are frequently used to separate higher frequency components in order to detect tremor, whereas stride time is used to characterize gait impairments \cite{mahadevan2020development}. 
%While the specifics of feature engineering and dataset generation will be covered in Section III, it is crucial to remember that these processes are essential for converting IMU signals into biomarkers with clinical significance.

\subsection{Multi-modal data acquisition via multiple sensors}

Numerous studies have looked into multimodal approaches, which combine various sensor types to provide more thorough and ecologically valid PD assessments, in order to get around the drawbacks of single-sensor systems.

By combining force plates, microphotographic cameras, and a monocular camera, He et al. created a 3D detection system that was 93.3\% accurate in differentiating PD patients from healthy controls \cite{he_2024_a}. An in-home study extended this concept to the home setting by implementing a smart platform that integrated cameras, ambient sensors, and wearables worn on the wrist to track symptoms while engaging in daily activities \cite{Morgan2022inhome}. The reference standard was high-resolution video, while environmental and wearable data continuously recorded variations that were frequently overlooked in clinical evaluations. Nevertheless, privacy issues, especially with regard to video, and discomfort from prolonged wearable use were raised.

Accessibility has been the subject of other multimodal studies. The viability of recording speech and movement characteristics for telemedicine applications using phone and headset setups was shown by Zhan et al. and Lakovakis et al. \cite{zhan_2018_using}, \cite{Iakovakis2020}. Their solutions still had drawbacks, such as hardware variability and noise, and uncontrolled environments could degrade data quality.

In order to simultaneously record rigidity, weakness, and tremor, Li et al. proposed a multimodal glove that combined flex sensors, EMG electrodes, and vibration/tremor sensors \cite{Li2023glove}. The glove's potential for long-term use was constrained by comfort concerns and calibration complexity, despite its high reliability and close alignment with clinical observations.

When combined, these studies demonstrate the potential of mixed-sensor systems to overcome the drawbacks of separate modalities. Multimodal approaches can improve PD screening's accuracy, ecological validity, and robustness by combining complementary data streams, such as motion, force, speech, and electrophysiological signals. But there are still issues with striking a balance between scalability, privacy, comfort, and complexity. Therefore, mixed-sensor approaches are a significant step toward integrated, practical monitoring solutions, acting as a link between thorough home-based care and controlled clinical evaluations.

\begin{figure}
    \centering
    \includegraphics[width=1\linewidth]{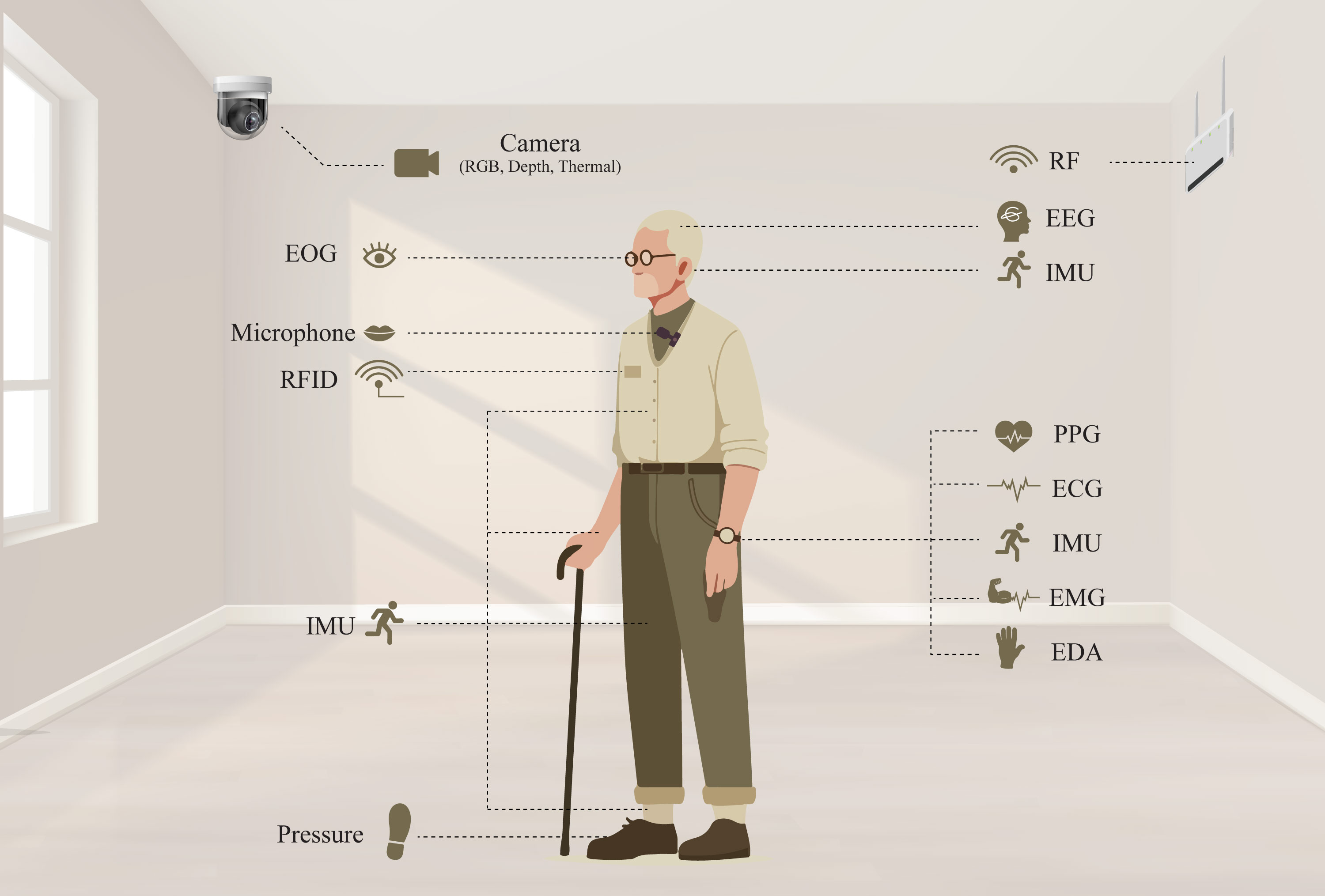}
    \caption{Wearables, IoT and IoMT sensors used for continuous monitoring of motor and non-motor symptoms of PD patients.}
    \label{fig:pd_sensors}
\end{figure}

Fig. \ref{fig:pd_sensors} provides a visual depiction of various wearables, IoT and IoMT sensors that researchers have used for PD research.

\subsection{Senors-based Datasets for PD Research}
Table IV in the Appendix summarizes a large number of studies/datasets that utilize a number of sensing modalities for Parkinson's disease research. Given the challenges of recruitment, ethics, and follow-up in clinical populations, it is not surprising that the majority of these datasets continue to be relatively small. More reliable group comparisons are made possible by the largest cohorts, such as gait datasets with almost 200 PD patients and more than 100 controls. The wealth of video-verified annotations is the strength of home-based freezing-of-gait studies, which, on the other hand, may only involve 20 participants. Most studies sit somewhere in between, and their findings should be considered in light of these sample sizes. 
A variety of methods are employed in the validation process. 10-fold splits are frequently used on EEG data, and K-fold cross-validation is frequently used. While some older studies use Monte-Carlo resampling, leave-one-subject-out (LOSO) is particularly useful for subject-independent testing, such as in at-home freezing of gait (FoG) detection. When dealing with small samples, repeated or k-fold runs help stabilize the results, and splits made at the subject level (as opposed to the window level) typically provide a more realistic view of model performance. 
Labeling conventions also vary. Some studies make use of broad clinical categories like UPDRS subscores, Hoehn \& Yahr staging, or PD versus healthy controls. Some even go so far as to annotate events in great detail, such as indicating the precise beginning and ending of a freezing episode. While home-based work mainly depends on challenging manual video labeling to secure accurate ground truth, webcam and sensor studies frequently link their extracted kinematics back to UPDRS ratings.
The majority of papers still rely on conventional machine learning techniques for modeling. After feature selection, SVMs, logistic regression, random forests, k-NN, and decision trees are frequently used on engineered time- and frequency-domain features. A number of studies use IMU data to test deeper models like CNN-LSTMs, but the simpler classical pipelines frequently outperform them, especially when subject-wise splits are used. 

Lastly, these works have different objectives. Some studies focus on symptom or event detection (identifying freezing episodes in day-to-day life), screening or diagnosis (differentiating PD from healthy controls), and disease severity monitoring (predicting UPDRS scores, evaluating EEG responses to cognitive load, or tracking changes in gait over time). The emphasis on early detection, before obvious motor signs manifest, is still comparatively uncommon, but it is a significant opportunity for future datasets that integrate subject-independent validation techniques, thorough clinical labels, and naturalistic sensing.

% ===== WHERE THE TABLE GOES =====

%\onecolumn
% temporarily make margins zero (or tiny) so the text block == page width
%\newgeometry{left=0.3cm,right=0.3cm,top=2cm,bottom=2cm}  % adjust top/bottom as you like

% restore your normal document margins/layout
%\restoregeometry
%\twocolumn

\section{Research Pillar II: Robotic Technologies for Parkinson's Disease Care}

This section summarizes the second set of research literature that focuses on development of various types of robots for PD patients under a wide range of usecase scenarios.

Robotic therapies have the potential to greatly enhance dexterity, strength, coordination, and motor function\cite{Banyai2024}. These systems have emerged as a highly effective therapeutic approach due to their ability to promote neuroplasticity through focused, repetitive, and high-intensity training\cite{Majeed2025}.  Although the emphasis of this section is on how robotics can improve motor function in people with PD, it's crucial to recognize their wider and more important role in the healthcare system. As evidence of their adaptability and expanding influence in contemporary medicine, robotic technologies are being widely used in fields like surgical assistance, rehabilitation, patient monitoring, medication delivery, and elder care\cite{Thimmaraju2025,Mendoza2025,mcgibbon2024exercising}. Within this broad spectrum, the applications most relevant to Parkinson’s disease rehabilitation fall into three main categories: robot-assisted rehabilitation systems, socially assistive robots, and integrated platforms that combine robotics with virtual or augmented reality.

\subsection{Robot-assisted rehabilitation systems (RARS)} 
RARS are a class of electromechanical systems designed to support, guide, or resist a patient's movements during rehabilitation. They aim to improve motor recovery after neurological or musculoskeletal disorders by key principles such as repetivity, task specificity, and real time feedback. RARS have shown significant improvements in motor symptoms, lower limb motor function, fatigue, and balance confidence in PD patients \cite{Tao2024}. These systems come with settings that are tailored to specific motor functions and rehabilitation goals.

RARS can be categorized based on their mechanical configuration into the following: upper-limb systems, lower-limb systems, Exoskeletons, and End-Effectors. The first type targets the rehabilitation of the shoulder, elbow, wrist, and hand, it aims to improve the fine motor control and grasping functions, while the second type is used to support the gait training, balance, and postural stability. Exoskeletons systems are wearable devices that move joints directly. On the other hand, End-effector devices move the hand/foot from the distal end, letting joints move indirectly. 
For PD patients, RARS have demonstrated positive results for several symptoms and motor impairments. For example, they could help increase movement amplitude and speed to mitigate bradykinesia, or enable passive stretching and assisted movement to alleviate rigidity and stiffness\cite{jiang2024effect, picelli2014robot}. Moreover, RARS-assisted rehabilitation offers repetition and intensity, which is crucial for fostering neuroplasticity and functional recovery. However, to enhance the outcomes of RARS-based rehabilitation for PD patients, it is essential to consider not only the intensity but also the duration and long-term continuity\cite{Tao2024}. 

In order to understand how RARS actually deliver assistance, it is imperative to understand their fundamental components.
The three fundamental components of the majority of wearable lower-limb rehabilitation systems are control, sensing, and actuation \cite{baud2021review}. Actuation usually involves (i) pneumatics, which offer high torque-to-weight and natural compliance but may require off-board air; (ii) compliant electric actuators, series elastic actuators (SEAs) and variable stiffness actuators (VSAs), which increase safety and shock isolation but come with additional mass and complexity; or (iii) direct-drive motors, which transmit force via Bowden cables in soft suits, keeping the hardware lightweight and close to the ankle \cite{sanchez2019compliant}. Multimodal sensing includes IMUs/encoders estimating joint kinematics and gait phase, force-sensitive resistor (FSR)/ground reaction force (GRF) sensors beneath the foot detecting heel-strike and toe-off, and occasionally EMG anticipating user intent \cite{manna2023optimal}. Simply put, the controller determines how much assistance to provide based on the user's position in the gait cycle, the actuators provide the assistive force, and sensors determine the user's location.

Mid- and high-level controllers typically use event-based (finite-state) strategies. When sensors detect heel-strike or toe-off, the controller transitions between gait phases (stance, swing) and selects a phase-specific assistance profile—for example, a brief torque pulse at push-off or open-loop support during stance. Position control can track a reference trajectory (or mirror the contralateral limb) when guidance is needed. %Beyond control strategies, compare robotic designs systematically, weighing complexity against practicality.

A recent review proposes a quantitative index to compare exoskeleton designs using four ingredients—assisted limbs (L), a degree-of-freedom proxy (D), sensor count (S), and actuator count (A), less weight given to the sensor, actuator comes next, then the freedom and assisted limbs contribute to more than half of the weights \cite{mallah2025systematic}.
Lower complexity means lighter, safer, easier to regulate, and more practical robotic system. 
Higher complexity indicates more capable, but heavier and harder system to use.

\subsection{Socially-assistive robots (SARs)} 
SARs place more emphasis on social and emotional support than robot-assisted rehabilitation, which is more concerned with physical movement. It focusses on motivational, emotional, and cognitive support. Since PD involves cognitive and psychosocial impairments, SARs could enhance the quality of life for individuals with Parkinson disease (IwPD). SARs can interact with users through verbal interaction, gesture recognition, and can adapt to individuals' needs or preferences by personalization algorithms.  In addition, they can monitor behavior and provide prompts or reminders. Also, they can provide encouragement for cognitive or physical exercise. Several studies, as well as clinicians, expressed positive impression toward integrating SARs within PD cares. In particular, SARs for IwPD were well received by clinicians when utilized in the patient's home to address motor, communicative, emotional, and cognitive demands, particularly for practice and assistance with everyday life tasks\cite{ResearchGate2025}.
Additionally, according to a study, a new socially helpful robotic platform shows promising outcomes: Positive engagement and motivation to use the platform in the short term were demonstrated during a feasibility test with 16 IwPD. In addition to incorporating interactive gamified elements, the platform is novel and unique in its targeted approach of using SARs for IwPD\cite{Raz2023}.

\subsection{Augmented and virtual reality-empowered robotic systems}
Looking ahead, the integration of robotic systems, augmented reality (AR), and virtual reality (VR) represents the most productive application and warrants promotion\cite{Tao2024}. This could highlight the strengths of physical robotics and virtual environments to create an engaging scenario. To elaborate, several studies have already started exploring this integration. For example, a recent study proposed a powered wheelchair equipped with a joystick and head-mounted VR display, and with this setup, with some additional components, the patient utilizes the joystick to navigate 20-meter virtual route, and then takes a 90 degree turn. This allows the system to capture data related to route deviation and movement velocity, which helps in determining the severity of tremor and motor control\cite{meyer2020mixed}. Such integrative platforms have the potential to deliver personalized care, especially when combined with sensor-based feedback mechanisms. Overall, the concept involves integrating machine learning, augmented reality, wearables, and soft robotics to increase the quality of life of IwPD through well-designed technology\cite{Mazumdar2025}.When combined, these various systems demonstrate the wide range of robotic innovation for PD, from social interaction to immersive hybrid environments to physical training, and they point to the possibility of more engaging and customized rehabilitation. 

\begin{figure}
    \centering
    \includegraphics[width=\linewidth]{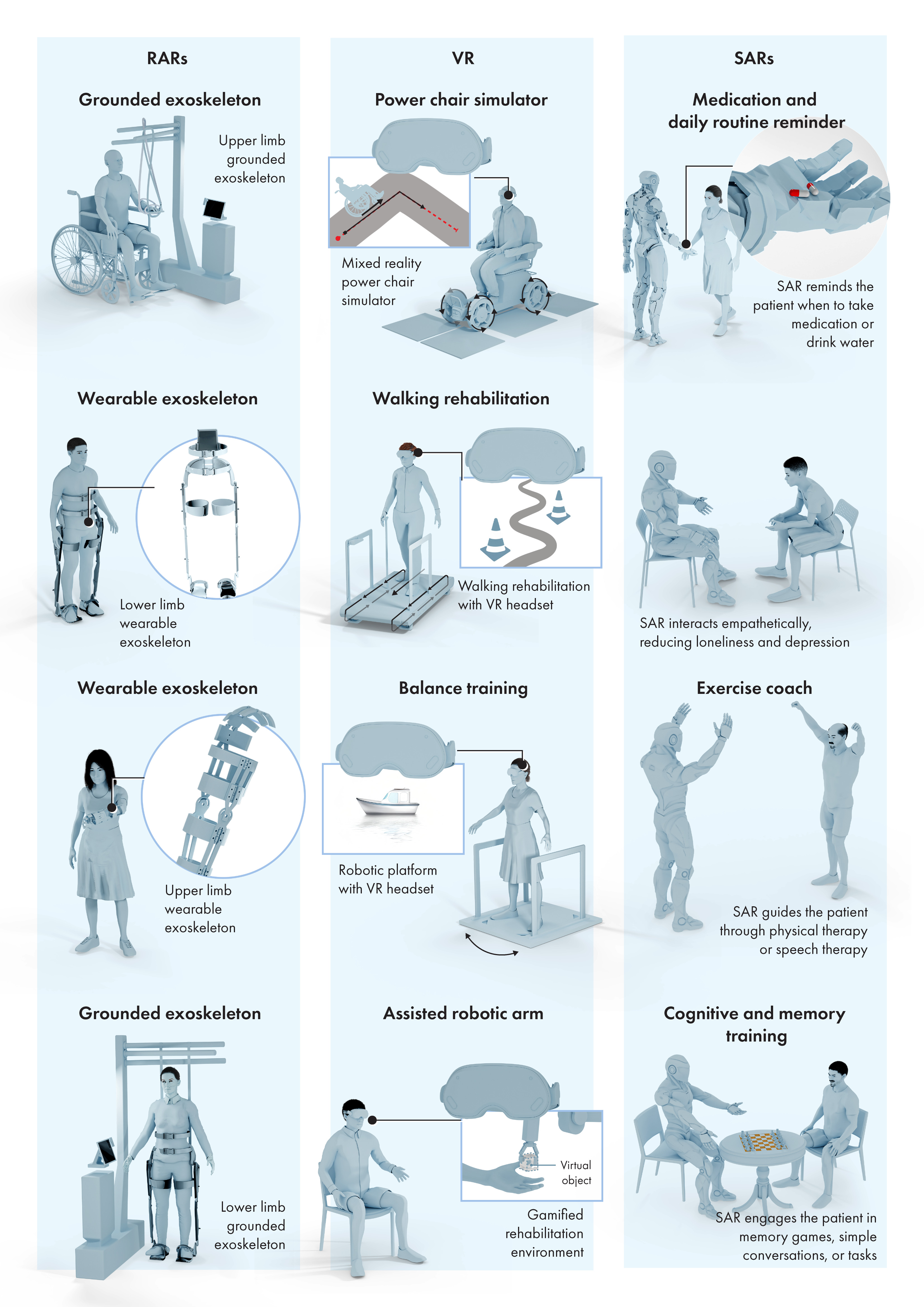}
    \caption{ RARS, SARs and AR/VR-empowered robotic systems hold the promise for rehabilitation of IwPD under various different scenarios. }
    \label{fig:robots_types_and_usecase_scenarios}
\end{figure}

Fig. \ref{fig:robots_types_and_usecase_scenarios} presents a visual summary of three kinds of robotic systems, i.e., RARS, SARs, AR/VR-empowered robots (and their sub-types) that could be used for care and rehabilitation of IwPD under various different scenarios.

\subsection{Comparison with traditional therapy methods}
It's also critical to compare these new robotic techniques to more traditional rehabilitation techniques in order to put them in perspective. There are several conventional methods for IwPD, such as physiotherapy (PT), occupational therapy (OT), and speech therapy (ST). PT aims to improve gait and flexibility through resistance and balance exercises. OT focuses on enhancing specific daily tasks by task adaptation, while ST targets speech clarity\cite{zotaj2024effectiveness,wood2022occupational,chang2023feasibility}. While these methods are still essential and widely used, robot assisted systems for training could be more effective in several areas.

%\begingroup
%\setlength{\belowcaptionskip}{0pt}     % shrink gap under the caption
%\setlength{\textfloatsep}{6pt}         % gap between the float and text (top/bottom placements)
%\addtocounter{table}{-1} 

\begin{table*}[!ht]
  \centering
  % first column fixed; other two are flexible ragged-right
  \begin{tabularx}{\linewidth}{|p{2cm}|X|X|}
    \hline
    \textbf{Comparison area} & \textbf{Conventional rehabilitation} & \textbf{Robotic-assisted rehabilitation} \\
    \hline
    Repetition & Determined according to therapist scheduling & Can deliver high-frequency sessions. \\ \hline
    Feedback & Depends on therapist observation & Real-time feedback depending on the sensors and algorithms used. \\ \hline
    Motivation & Relies on harmonization between patient and therapist & May include interactive elements that promote engagement (e.g., SARs). \\ \hline
    Personalization & Based on therapist judgment; flexible and intuitive & Via sensors and ML, but less aware of the human condition (for now). \\ \hline
    Data collection & Manual documentation & Automated data collection. \\ \hline
    Cost & Low cost, but requires clinician time & Higher cost but improved accessibility. \\ \hline
    Human/emotional interaction & Strong human connection & Limited emotional intelligence (for now). \\ \hline
  \end{tabularx}

  % add a little breathing room above the caption (local to this table)
  %\vspace{0.4\baselineskip}

  \caption{Comparison between conventional and robot-assisted rehabilitation for PD care \cite{Spandidos2025}.}
  \label{tab:training_comparison}
\end{table*}
 
Table \ref{tab:training_comparison} compares conventional and robot-assisted rehabilitation in terms of repetition, feedback, motivation, personalization, data collection and monitoring, cost and accessibility, human and emotional interaction.
Additionally, a study that compared the effects of treadmill training and robot-assisted gait training (RAGT) quantitatively in PD patients found that patients who received robotic training had better gait kinematics, especially in the frontal plane at the pelvic and hip joint level \cite{Spandidos2025}. 
Similarly, an early trial by Picelli et al. revealed a statistically significant difference favoring robotic stepper training (in contrast to physiotherapy with active joint mobilization). For up to one month, the beneficial impact persisted\cite{Spandidos2025}. 
These findings suggest that robotic systems offer promising biomechanical and functional benefits, and they could play an important role in the future of PD rehabilitation.

\section{What is on the Horizon? Human-AI-Robot Collaboration and AI Agents-based Digital Twin for PD Care}

%Over the past decade, AI-enhanced robotic surgery has emerged as a paradigm shift in modern medicine, offering substantial gains in precision, efficiency, and patient outcomes. A 2025 meta analysis synthesizing 25 peer reviewed studies across specialties-including urology, orthopedics, neurosurgery, and head and neck oncology-found that AI driven robotic procedures reduce operative time by approximately 25\% and intraoperative complications by 30\%, while improving surgical targeting accuracy by 40\% and lowering postoperative recovery times by around 15\% \cite{wah2025rise}. Recent reviews in oncology specifically highlight innovations that improve surgical safety and care personalization, including adaptive motion control, real-time navigation, image-free palpation, and predictive tumor resection \cite{wah2025ai}.

\subsection{Opportunities}

\textbf{Humanoid Robots for PD care:}
Recently, humanoid robots such as Unitree G1 are starting to function actively as all-purpose clinical assistants. Unitree G1 can now perform seven different medical interventions, including auscultation, emergency ventilation, palpation, and ultrasound-guided needle injections, thanks to a recent exploratory study that created a bimanual teleoperation system that shows promise in real-world trials\cite{atar2025humanoids}. Despite issues with force control and sensor fidelity, the platform demonstrated flexibility and clinical accuracy in critical tasks (such as ventilation and needle injections), achieving quantitatively promising results despite not being fully autonomous\cite{atar2025humanoids}. This puts humanoid robots in a position to be useful support systems in remote or understaffed care settings as well as high-burden hospital settings. 
Shanghai RobotGym's Qijia Q1 has a novel and useful design that combines a wheelchair and a human. The elderly can benefit from its assistance with wake-up calls, prescription reminders, breakfast preparation, entertainment, and outdoor activities as a daily companion and aide.\cite{VCBeat_QijiaQ1_2023}.

\textbf{Foundation Models for Robotic Systems:}
Recent advances in robotics foundation models, aka vision-language-action (VLA) models, aim to replace narrow, task-specific policies with generalist policies capable of transferring skills across tasks, objects, and embodiments. Unlike traditional modular pipelines, VLAs integrate robot action heads trained on task demonstrations with internet-scale vision and language pretraining, thereby enabling open-vocabulary perception, zero-/few-shot task adaptation, and instruction-following capabilities (e.g., pick up the blue cup on the left). Early examples such as PaLM-E and RT-2 have demonstrated how web-scale semantics can be embedded into robot control to improve generalization outside the training distribution \cite{zitkovich2023rt,driess2023palm}.

The current design space is illustrated by representative open and closed systems. OpenVLA, a 7B-parameter model pretrained on approximately 970k episodes, supports multi-robot control and efficient adaptation via parameter-efficient fine-tuning (PEFT). Octo introduces a flexible transformer-based policy, pretrained on ~800k trajectories, that accepts either natural language commands or goal images, supporting rapid adaptation to novel sensors and action spaces. Similarly, RoboCat exemplifies self-improving agents that iteratively expand their competence by collecting additional demonstrations ($\approx$$10^{2}$-$10^{3}$) using the model itself. Collectively, these systems mark a transition from single-robot, single-task pipelines to reusable, skill-centric controllers \cite{kim2024openvla}.

The diversity and scale of robot datasets have emerged as key enablers for such generalist models. The Open X-Embodiment (OXE) project aggregates cross-robot and cross-institution datasets into standardized formats to facilitate cross-embodiment transfer. BridgeData V2 provides $>$60k trajectories across 24 environments on cost-effective hardware, supporting reproducibility. Meanwhile, DROID expands the training distribution through in-the-wild demonstrations, comprising roughly 76k trajectories (~350 hours) spanning 564 scenes and 84-86 tasks. Pretraining on such large, heterogeneous datasets improves robustness in novel settings, e.g., home and clinical settings where distribution shifts are common \cite{vuong2023open}.

\textit{Relevance to PD care:} Many activities of daily living (ADLs) critical for individuals with PD---such as organizing medication, transporting items, and operating household objects---require long-horizon mobile manipulation. Recent advances in cost-effective bimanual, whole-body teleoperation platforms, such as Mobile-ALOHA, provide scalable mechanisms for collecting demonstrations of such ADLs. VLA models pretrained on large-scale datasets (e.g., OXE, Bridge, DROID) can be adapted through PEFT/LoRA to support PD-specific contexts. A practical pipeline would involve pretraining on broad multi-robot datasets, selecting an open VLA (e.g., OpenVLA or Octo) for instruction-following and manipulation skills, and refining these models on clinician-designed task sets recorded in controlled clinical or mock-home environments. Iterative refinement through AutoRT-style data collection could further personalize the system.

This framework suggests a closed-loop robotic care system for PD management, where an LLM/agentic layer provides explanation, personalization, and safety oversight; wearable sensing modalities (e.g., IMUs) and SARs deliver contextual and therapeutic support; and the VLA governs the execution and continual learning of assistive motor tasks \cite{li2024llava,kim2024openvla}.

\textbf{LLM and Agentic AI Frameworks for PD Care:}
The integration of large language models, conversational AI, and agentic AI systems with robotic and sensing technologies, such as IMUs and other wearable devices, holds the potential to reshape early diagnosis, prognosis, and daily management of PD \cite{yao2024survey}, \cite{mohammed2023innovative}. Unlike traditional AI systems that rely heavily on predefined instructions and human oversight, agentic AI frameworks incorporate reasoning, adaptability, and goal-directed behavior into clinical decision-making. Such systems can perceive their environment through multimodal sensory inputs, process contextual information, and autonomously determine optimal actions to achieve specific healthcare objectives. This combination of adaptability and autonomy positions agentic AI as a promising paradigm for PD care, complementing the linguistic and interpretive capabilities of LLMs.

Several recent studies demonstrate the application of LLMs and agentic AI in PD treatment and management. Zhang et al. \cite{zhang2025leveraging} proposed an LLM-based decision-making framework that outperformed reinforcement learning strategies and even physician-guided baselines. The LLM demonstrated the ability to interpret patient symptom descriptions alongside clinical guidelines, thereby supporting its potential role in augmenting treatment planning. Similarly, Cardenas et al. \cite{cardenas2024autohealth} developed a system that integrates smartwatch-derived IMU data with LLM capabilities. In their design, raw sensor data is transmitted to a smartphone, processed, and subsequently interfaced with an LLM-powered chatbot. This system is intended to function as a self-learning agent that assimilates patient-specific healthcare records and longitudinal histories to deliver personalized care strategies. In another study, Mohammed et al. \cite{mohammed2023innovative} employed LLM-based techniques to analyze dyskinesia data, enabling AI-driven severity assessments and automated recommendations for patient management.

Collectively, these works illustrate the emerging role of LLMs and agentic AI in enhancing PD care by enabling personalized, context-aware, and interactive clinical support systems. Table \ref{tab:LLM} provides a comparative summary of these studies, highlighting their methodologies, sensor integration strategies, and key contributions toward advancing AI-driven PD management.

\begin{table*}[t]
\centering
\begin{adjustbox}{width=\textwidth}
    \begin{tabular}{|p{2cm}|p{4cm}|p{9cm}|}
    \hline
        \textbf{Reference} & \textbf{Used technology} & \textbf{Modality} \\ \hline
        \cite{zhang2025leveraging}& LLM (Monte Carlo Tree Search (MCTS), Similar Case Retrieval, chain of thought (CoT)) & Using MCTS to generate guided treatment paths for Parkinson disease patients, with decisions informed by patient measurements, a large language model, and retrieval of similar historical cases. \\ \hline
        \cite{macedo2024conversational}& AI conversational agent & AI agent (chatbot) integrated into a mobile application, which will receive voice or text inputs from the user, and respond accordingly offering advice, exercise demonstrations, motivational support, and other features. \\ \hline
        \cite{cardenas2024autohealth}& LLM (NLP chatbot) + IMU (Accelerometers) & An accelerometer and microphone detect movement and speech data, which is sent via Bluetooth to a smartphone for PD detection. A chatbot supports user interaction. \\ \hline
        \cite{mohammed2023innovative} & LLM (LSTM-based time series) + IMU (Accelerometers, gyroscopes, and pulse sensors) & A palm-worn device integrates accelerometers, gyroscopes, and pulse sensors to measure data. LLM is used to assess PD severity. \\ \hline
        \cite{crawford2025linguistic} & LLM (BERT, XLNet, GPT-2, text-embedding-ada-002, text-embedding-3-small/large) & Natural speech is converted to text, embedded using LLMs to extract linguistic features, which are input to a classifier for PD detection. \\ \hline
        \cite{varkuti2024conversational} & LLM (GPT-4) & Provides patients with precise and up-to-date information about Parkinson disease. \\ \hline
    \end{tabular}

\end{adjustbox}
\caption{Related works that discuss the use of LLMs alone, or LLMs integrated with wearable sensors, for PD care.}
\label{tab:LLM}
\end{table*}

\textbf{Digital Twin for PD Care:}
Recently, there have been a few efforts for development of digital twins (DTs) that reliably replicate the traits, behaviors, and interactions of IwPD \cite{awasthi2025exploring}. A PD digital twin is meant to be as a bi-directional and interactive framework, capable of receiving multimodal data from patients and providing predictions, recommendations, and feedback to improve treatment personalization \cite{ashraf2024digital}. For example, Rahman et al. \cite{rahman2024user} proposed a video-based approach to assess PD severity, supported by an integrated automated quality control tool that delivers real-time feedback on video quality and filming conditions. In parallel, LLMs have been considered for caregiver and family support; however, their effectiveness requires carefully constructed prompts that are comprehensive, task-specific, and unambiguous, to minimize hallucinations \cite{rahman2024user}.

The foundation of the PD DT lies in the real-time acquisition of physiological and behavioral data via wearables, sensors such as inertial measurement units (IMUs), and other monitoring instruments. This data is then utilized to construct a virtual representation of the IwPD, enabling continuous monitoring of disease states and progression. For instance, Qin et al. \cite{qin2024design} proposed a multimodal DT system that integrates synchronized sensory inputs---including speech, hand and upper-body motion video---with static clinical data. The system is supported by three server-side models tailored for rehabilitation management: hand-tremor recognition (98.70\%), speech recognition (95.68\%), and tremor-severity evaluation (75.2\%) using TS-STGCN, a graph-based deep learning architecture specialized in skeleton motion analysis. Similarly, Abirami and Kartikeyan \cite{abirami2023digital} developed a cloud-based DT that employs voice recordings captured via smartphones to build patient-specific virtual profiles. While the system is designed to accommodate five severity levels of PD, their evaluation was restricted to binary PD-versus-healthy voice classification. Using an Optimized Fuzzy k-Nearest Neighbor classifier, their approach achieved accuracies of 97.95\% and 91.48\% on two benchmark datasets, with correspondingly high F1 and MCC scores, indicating strong performance in dataset-level PD detection tasks.

Digital twins for PD management can also benefit from integration with other emerging technologies \cite{abd2024digital}. For instance, Sharafian et al. \cite{sharafian2025effects} demonstrated that a smartphone-organized, IMU-driven vibrotactile cueing system significantly improved stride length ($\approx$14\%) and gait speed ($\approx$18\%) in older adults during a single session, with only modest increases in reaction time ($\approx$27-74 ms). Such results provide promising evidence for the feasibility of individualized, closed-loop, home-based interventions that may be adapted for PD rehabilitation.

\subsection{Challenges}

\textbf{Data acquisition:}
One of the most fundamental barriers is the difficulty of gathering high-quality data from real PD patients. Many studies still work with small cohorts and short monitoring periods, limiting the strength and generalizability of their findings. Recruiting large, diverse participant groups, especially for long-term follow-up, demands significant time, funding, and coordination. Busy clinics frequently lack the time necessary for thorough evaluations, which results in datasets that are inconsistent or lacking \cite{alves2025stakeholder}. Continuous data collection is difficult even outside of the clinic since participants may forget to wear their devices, experience technological difficulties, or eventually stop participating.

\textbf{Data labeling and annotation:}
Another difficulty is that this data must be labeled before it can be used to train algorithms. For example, to detect episodes of FoG, experts still need to watch hours of video and carefully record the exact frame at which each episode begins and ends \cite{chen2025freezing,yang2024freezing}. This process is costly, time-consuming, and necessitates specialized knowledge. Another degree of uncertainty is introduced by the possibility that various reviewers will view the same occurrences slightly differently.

\textbf{Multi-modal sensing:}
Another disadvantage is that no single sensing method can adequately capture the variety of PD symptoms. Wearable IMUs are excellent at detecting abnormalities in gait and motor indications, but they don't tell us much about non-motor symptoms like mood changes, speech abnormalities, or cognitive decline. A more complete picture can be obtained by combining different types of sensors, such as physiological trackers, wearables, and environmental monitors, but doing so also creates new challenges for data fusion, device design, and analysis \cite{oyama2023analytical }.

\textbf{Rare symptoms:}
The rarity of certain symptoms presents another difficulty from the perspective of machine learning. For instance, FOG might only occur once during a full recording day, which would seriously throw off the collection's balance. Models trained on such data run the risk of overfitting to the majority class (normal walking), and simple metrics like accuracy or ROC curves can give a false sense of performance. More complex evaluation methods, like precision-recall curves or re-sampling techniques, are needed to assess these algorithms objectively \cite{mlcontest2024}.

\textbf{Trustworthiness and privacy of PD care:}
Last but not least, even when technology works well, patient comfort and trust are important determinants of acceptance. Continuous home monitoring may seem intrusive, particularly if cameras or microphones are being used. Many participants ask for clear boundaries, such as prohibiting recording in private spaces or during particular hours. Giving patients a great deal of control over the timing and mode of monitoring is essential to easing these concerns, as is having robust data security and transparent consent procedures \cite{ radanliev2025privacy}.

\textbf{Other challenges:}
Some other challenges include: data governance/consent, privacy, and on-device inference (which is crucial in patient homes). Also, objective evaluation beyond short-horizon tabletop tasks is limited; simulation-to-reality bias still exists. Thus, there is a need to do focused clinical translation research in PD populations (e.g., caregiver-in-the-loop workflows, rehabilitation coaching, and assisted daily living-focused assistance of IwPD) \cite{ xu2024survey}. Finally, standardization efforts for AI-powered digital twin robots (from patient safety, data privacy perspectives) need to be initiated.

\subsection{Authors' vision}

%The management of Parkinson's disease is poised for a paradigm shift, moving from episodic, reactive interventions to a continuous and proactive model through the integration of AI and LLM-powered robotic assistants. We envision a future where these agents serve as intelligent partners within the care ecosystem, augmenting human capabilities rather than replacing them. 

We envision a futuristic multi-agent paradigm for PD care in which wearables, medical IoT devices, ambient sensors, and socially assistive or rehabilitative robots converge into a seamless, intelligent, and personalized management ecosystem. Existing research on PD care has progressed substantially along two parallel pillars: (i) multimodal sensing technologies such as IMUs, EMG, acoustic analysis, force/pressure insoles, cameras, and RF-based passive monitoring systems for detection and monitoring motor and non-motor symptoms of IwPD; and (ii) robotic and immersive technologies, including exoskeleton-based rehabilitation robots, socially assistive robots, and AR/VR platforms. While each has demonstrated clinical utility in isolation, the next transformative step lies in their integration into a unified, AI-driven framework.

At the heart of this vision is the Parkinson's patient digital twin (PPDT), a continuously updated, high-fidelity model of the patient's physiological, behavioral, and clinical state. In this scenario, three key stakeholders---the patient, caregiver, and a physically embodied humanoid robot---interact locally, while a fourth stakeholder---the treating physician---connects remotely via secure telemedicine channels. Multimodal data streams from patient wearables (e.g., gait kinematics, tremor severity, speech prosody), caregiver context (e.g., assistance patterns), and robotic platforms (e.g., force interactions during therapy) are pre-processed on embedded AI systems within the robot, enabling privacy-preserving, real-time inference.

The PPDT is powered by a digital twin AI agent, a modular intelligence stack with several key components. A foundational multimodal model, trained on large-scale wearables datasets, provides generalization to unseen patients, while a multimodal fusion engine integrates heterogeneous data to infer PD-specific biomarkers such as UPDRS scores or freezing-of-gait episodes. A reinforcement learning with human feedback (RLHF) module incorporates structured and unstructured feedback from patients, caregivers, and physicians, closing the loop between sensing, interpretation, and intervention. A retrieval-augmented generation (RAG) subsystem allows the agent to dynamically access PD guidelines and medical literature, while large language and acoustic foundation models enable natural, empathetic interaction with patients and caregivers. Finally, meta-learning and continual learning components ensure adaptability to disease progression and intra-patient variability.

This closed-loop architecture facilitates continuous evaluation, feedback, and tailored interventions. Patients and caregivers receive simplified insights through mobile applications, while physicians access clinically relevant dashboards. Privacy-preserving frameworks such as federated learning and blockchain-based encryption ensure secure data handling and compliance with international standards. Beyond monitoring and decision support, the integration of socially assistive robots provides physical and emotional engagement, first-aid capabilities, and even emergency interventions (e.g., automatic SOS alerts after falls).

\begin{figure}
    \centering
    \includegraphics[width=\linewidth]{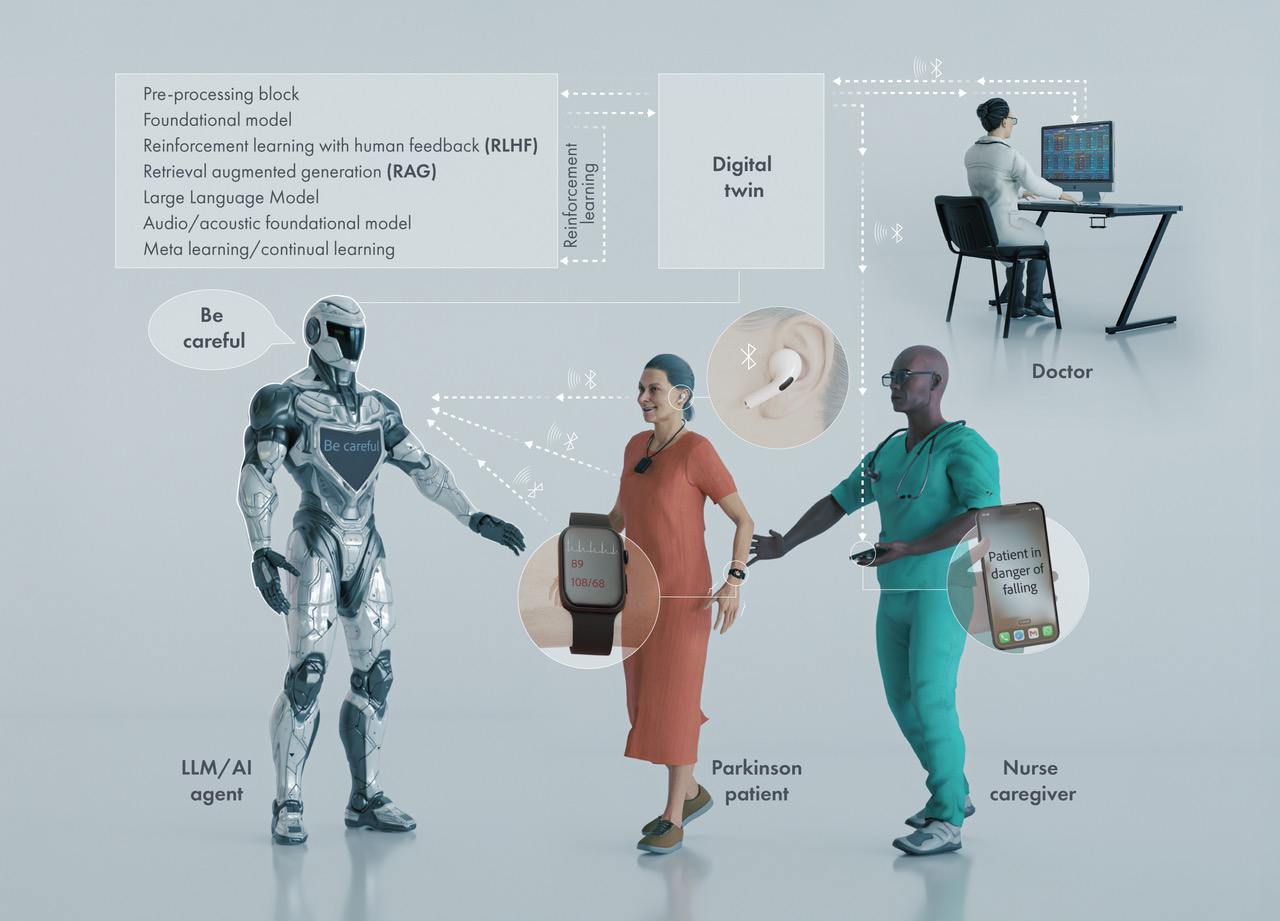}
    \caption{Authors' vision: continuous monitoring and personalized care for Parkinson’s disease via agentic AI-powered robotic digital twins.}
    \label{fig:vision}
\end{figure}

By maintaining a live and adaptive digital twin, this framework could revolutionize PD management through early detection of symptom fluctuations, personalized rehabilitation and medication guidance, and enhanced caregiver support. 
Figure \ref{fig:vision} depicts the envisioned PPDT system and outlines each component within a cohesive PD care framework.

Let's further illustrate the strength of an Agentic-AI-based robotic digital twin system by comparing it with the traditional, human-centric model. The traditional caregiver based support system is characterized by subjective assessment and latent response; a patient's fall relies solely on a caregiver's manual evaluation, while subtle symptom decline like speech softening may go unaddressed for weeks until a clinical visit. In contrast, the envisioned PPDT support system enables assisted independence, where an inertial measurement unit can autonomously detect a fall, classify its severity, and initiate stabilization and an emergency alert, or where gait deterioration triggers the Agentic-AI-based robot to propose personalized rehabilitation games based on historical data.

The optimal framework, however, is a hybrid collaborative model that highly depends on the relationship between the patient, caregiver, and robot. In this paradigm, the system's continuous monitoring and precision complement the caregiver's empathetic judgment. For instance, the Agentic-AI-based robot can act as a first responder in an emergency, physically stabilizing a patient and alerting the physician, while simultaneously notifying the caregiver for verification and oversight. In daily management, the PPDT system's ability to detect early dyskinesia allows it to prompt patient exercises while notifying the caregiver with objective data to support timely medication adjustments. This collaborative approach transforms care from being reactive to preemptive, enhancing patient safety and independence while empowering caregivers with data-driven insights, ultimately elevating the standard of support through intelligent automation.

On the flip side, we note that standardization efforts for AI-powered digital twin robots (from patient safety, data privacy perspectives) also need to be initiated by the key stakeholders from academia, industry, public health, and hospitals.

Last but not the least, the modular design of PPDT system allows to utilize AI tools such as transfer learning, knowledge distillation, fine-tuning and domain adaptation for its utilization for other chronic conditions such as Alzheimer’s disease \cite{cardenas2024autohealth} and dementia, thereby broadening its clinical impact.

\section{Conclusion}
This survey examined a wide spectrum of technologies for PD management, from noninvasive sensing (IMUs, pressure insoles, EEG/EMG, speech, vision, and radio) and the corresponding public and private datasets, to robot-assisted rehabilitation, to emerging AI frameworks (large language models, agentic AI, and digital twins). We discovered that, across modalities, meticulous labeling and subject-independent evaluation are just as important as sensor selection: studies that report event-level annotations (e.g., clinician-verified FoG) and enforce LOSO/subject-wise splits produce more accurate performance estimates than window-randomized designs. Socially assistive robots provide additional cognitive-motivational support in the home, and rehabilitation robotics—both exoskeletons and end-effectors—show quantifiable improvements in gait kinematics and functional confidence when training is repetitive, task-specific, and feedback-rich.

When combined, these patterns generate a closed-loop PD care architecture in which digital or robotic interventions adjust in real time, AI models contextualize and forecast symptom dynamics, and multimodal sensors record daily variability. Then, based on the results, the next action is improved. Small, diverse cohorts; annotation cost and inconsistency; rare-event imbalance; and privacy, trust, and workload in real-world deployment are some of the practical limitations that have come up time and time again in the literature and must be taken into consideration in order to realize this vision. As a result, we list three priorities: (i) data and evaluation standards (transparent label provenance, precision-recall metrics for rare events, leakage checks, and subject-wise splits by default); (ii) privacy-by-design development (unified learning, on-device inference, and audited data governance) to gain the trust of patients and clinicians; (iii) human-in-the-loop AI that gives clinicians control, exposes uncertainty and reasoning, and uses LLMs/AI agents to suggest decisions.

Expanding valid datasets with strict labeling, benchmarking models under subject-independent protocols, combining sensing with adaptive, customized robotic training, and testing digital-twin and agentic workflows that safely close the loop are all clear steps in the near future. Together, these actions have the potential to transform today's promising prototypes into reliable, scalable systems that enhance the functionality and quality of life for those with Parkinson's disease.

\input{appendix}

\section*{Acknowledgements}
Ana Bigio, a scientific illustrator, produced the Figs. 3, 4. 

%\input{appendix}
% Citing all 28 references
\footnotesize{
\sloppy
\bibliographystyle{IEEEtran}
\bibliography{references}
}

\end{document}

%% file: appendix.tex
\section*{Appendix}

%%%%%%%%%%%%

\onecolumn

\begin{landscape}
\begin{longtable}{p{2.5cm}p{2.2cm}p{2cm}p{2.5cm}p{2.8cm}p{2.8cm}p{2.5cm}p{2cm}p{2.5cm}}
\caption{Sensors Based Datasets for PD Care.\label{tab:pd_camera_landscape}}\\
\toprule
\textbf{Work} & \textbf{Modality} & \textbf{Subjects} & \textbf{Data splitting strategy} & \textbf{Data labeling strategy} & \textbf{AI models used} & \textbf{Problem solved} & \textbf{Accuracy} & \textbf{Duration of recordings} \\
\midrule
\endfirsthead

\toprule
\textbf{Work} & \textbf{Modality} & \textbf{Subjects} & \textbf{Data splitting strategy} & \textbf{Data labeling strategy} & \textbf{AI models used} & \textbf{Problem solved} & \textbf{Accuracy} & \textbf{Duration of recordings} \\
\midrule
\endhead

\midrule
\multicolumn{9}{r}{\emph{Continued on next page}}\\
\bottomrule
\endfoot

\bottomrule
\endlastfoot

\cite{van2021camera} & Kinect V2 RGB-D & 19 PD + 8 HC & No ML train/validation/test split & PD/HC, ON/OFF; MDS-UPDRS-III; algorithmic walking-phase segmentation & No classifier or predictive ML model & Can a low-cost depth camera quantify PD gait and ON/OFF changes & No accuracy metric & 3 TUGs per state; $\sim$90 min between OFF$\rightarrow$ON; 30 Hz \\

\cite{wu2023kinect} & Azure Kinect RGB-D & 50 PD + 25 HC & No train/validation/test split & Binary PD vs HC labels & Kinect feature extraction + logistic regression & Diagnosis/screening: differentiate PD vs HC & 3-feature model: AUC 0.955 & 10 repetitions per side/hand, 1-min rest between tasks \\

\cite{monje2021remote} & Laptop webcam (MDS-UPDRS III bradykinesia tasks) & 22 PD + 20 HC & 4-fold cross-validation & MDS-UPDRS III scored off medication by 2 specialists & SSD + OpenPose; Logistic Regression, Naive Bayes, Random Forest & PD vs HC classification & ACC $\sim$0.33–0.83 & 12 s video clips; each task 3× per hand \\

\cite{lange2025computer} & Multi-camera RGB (6 Basler acA1920-155uc) & 15 (10 PD + 5 HC) & LOSO validation & UPDRS III scored by 3 neurologists & SVM, Random Forest, AdaBoost & UPDRS severity prediction + PD vs HC & Finger tapping: F1=0.69 (UPDRS), 0.87 (PD/HC); gait: 0.72 (UPDRS) & $\sim$1 min gait recording per subject; finger tapping in 10-tap windows \\

\cite{rusz2021speech} & Head-mounted condenser mic (speech) & 150 iRBD, 149 PD, 149 HC & Group discrimination via logistic regression & Clinical diagnoses & Least-squares linear models & Distinguish PD vs HC, PD vs iRBD, iRBD vs HC & AUC=0.80 (PD vs HC) & Reading 80–115 words + $\sim$90 s monologue \\

\cite{tsanas2009accurate} & Intel AHTD with head-mounted mic & 42 PD & 5923 phonations split 90/10 train/test & Ground truth UPDRS clinically assessed & LS, IRLS, Lasso regression & Weekly progression tracking & MAE $\approx$ 5.8 (motor UPDRS), $\sim$7.5 (total UPDRS) & Six phonations/day, max 30 s each \\

\cite{rusz2018smartphone} & Head-mounted mic + smartphone & 30 PD + 30 HC & Leave-one-out CV & PD, iRBD, HC by neurologist & Logistic regression + smartphone features & Early PD screening & AUC=0.85 (sens. 75\%, spec. 78.6\%) & Sustained /a/, rapid /pa-ta-ka/, 90 s monologue \\

\cite{molcho2023evaluation} & Single-channel forehead EEG (Neurosteer®) & 32 suspected PD + 20 HC & 1/3 withheld for test & F-DOPA PET positive/negative & PCA + GLMM/LMM & Early-diagnosis screening & Perfect classification of 14 F-DOPA+ and 6 F-DOPA- & $\sim$30 min session \\

\cite{aljalal2022parkinson} & Resting-state EEG (2 public datasets) & 31 (San Diego) + 54 (UNM) & 10-fold CV & PD vs HC, ON vs OFF & KNN, SVM, RF, LDA, QDA & Automated PD detection & Best off-PD vs HC $\approx$ 99.41\% accuracy & $\sim$2 min per subject (1 min eyes closed + 1 min eyes open) \\

\cite{lanzani2025methodological} & Review of EMG-based muscle-synergy studies & Primary studies & N/A & PD vs HC, ON vs OFF, Tremor vs voluntary & NMF, PCA & Map PD neuromuscular alterations & N/A & Up to 10 min walking, quiet standing 25–30 s \\

\cite{navita2025gait} & 16 in-shoe sensors & 166 (93 PD, 73 HC) & 70/30 split + 10-fold CV on train & Stage 1: PD vs Non PD; Stage 2: severity & RF, ensemble regressor & Screening + severity & PD detect: 97.5\%, Severity: 96.4\% & 306 recordings = 19320 gait cycles \\

\cite{Karamanis2024} & Pressure insole & 19 PD & NR (no ML split) & MDS-UPDRS III, axial, bradykinesia & No predictive model & Progression tracking & NR & 2-year follow-up \\

\cite{oday2022assessing} & IMUs & 7 PD (FOG) + 16 surveyed & LOSO CV & 2-s IMU windows labeled by majority rating & 1D CNN (2-layer) & FOG detection & AUROC 0.83 (lumbar+ankles) & 88.8 min total walking \\

\cite{ko2019measuring} & IMU (accel+gyro) & 11 PD (FOG) & LOSO CV & Windows labeled from video annotations & CNN + LSTM & FOG detection & F1=0.94, sens=0.91, spec=0.96 & $\sim$6 h IMU walking data \\

\cite{deepfogsensor2021} & IMUs (accel+gyro) & 13 PD & LOSO CV & Video-labeled FoG episodes & CNN & FoG detection & AUROC=0.94, sens=0.92, spec=0.96 & 7.1 h data over 135 trials \\

\cite{papadopoulos2020detecting} & Wrist-worn IMU & 45 PD & 10-fold CV & Tremor labeled by experts & Deep MIL + CNN features & Tremor detection & AUROC=0.902 & $\sim$436 h total \\

\cite{ullrich2023fall} & Lower-back IMU & 34 PD & LOSO CV & Fallers/non-fallers (self-report) & LR, SVM, RF, GB & Fall risk prediction & Acc=0.82 (GB) & 7 days continuous data/subject \\

\cite{delgado2025ankle} & Smartphone IMU & 55 (40 PD, 15 HC) & 5-fold CV & PD/control clinical diagnosis & CNN & PD classification & Acc=0.982 & 20–30 s per trial \\

\cite{salomon2024machine} & Wrist-worn IMU & 185 (PD + controls) & Train/test split + 5-fold CV on train & Clinical Dx + severity & Transformer models & PD detection + severity & AUROC=0.92 & 14 days/subject \\

\cite{cordillet2025detectfog} & 7 IMUs (thighs, fibulas, feet, lumbar) & 18 PD & LOSO CV & FoG onset/offset from video & Decision tree + Relief-F & FoG detection & N/A & Up to 18 trials/visit, 2 visits \\

\cite{mdpi2024weakly} & 12 activities (RAM, finger-to-nose etc.) & 83 (70 PD + 15 HC) & Leave-one-out CV & H–Y stages & MIL + k-means + LDA + XGBoost & Stage 4 classification & 73.48\% (aug), 71.08\% (no aug) & NR \\

\cite{souza2022public} & Shank IMU & 35 PD & N/A & FoG labeled from video & N/A & Public FoG dataset & N/A & 2 min per trial × 3 sessions \\

\cite{tsakanikas2023gait} & Insoles + 5 IMUs & 19 PD & NR & MDS-UPDRS gait ratings & SVM, RF, GB, AdaBoost & Binary classification & AUC=0.93, Acc$\approx$0.89 (SVM) & NR \\

\cite{adams2021realworld} & 5 wearable accelerometers & 17 PD + 17 HC & N/A & Tremor threshold from in-clinic video & N/A & Real-world PD activity quantification & N/A & Avg 45.4 h/subject \\

\cite{smartinsole2021} & Moticon smart insoles + IMU & 29 (13 HC, 9 elderly, 8 PD) & N/A & 12 ADL labels + gait events & N/A & Public elderly/PD gait dataset & N/A & N/A \\

\cite{varghese2024pads} & Two wrist-worn smartwatches & 469 (276 PD, 114 DD, 79 HC) & 5×5 nested CV (subject-wise) & Clinical Dx (PD/DD/HC) & SVM, CatBoost, NN, BOSS, XceptionTime & PD vs HC, PD vs DD classification & Bal. Acc=91.16\% (PD vs HC), 72.42\% (PD vs DD) & $\sim$15 min total \\

\cite{rodriguezmartin2017home} & Waist-worn accelerometer & 21 PD (FoG) & Leave-one-patient-out & FoG episodes labeled from video & RBF-SVM & Home FoG detection & Sens=88.1\%, Spec=80.1\%, GM=84.0\% & Two $\sim$20-min sessions (OFF/ON) \\

\cite{trabassi2022ml} & Lumbar IMU & 64 PD + 64 HC & 80/20 split + 10-fold CV & Clinically diagnosed PD vs HC & SVM, RF, DT, KNN, MLP & Binary classification & SVM Acc$\approx$0.86 & Corridor walk (duration NR) \\

%\cite{caramia2018classification} & Eight IMUs (15 m walk) & 25 PD + 25 HC & 5-fold CV ×100 & PD vs HC clinical labels & NB, LDA, k-NN, DT, SVM-linear/RBF & Binary PD vs HC classification & Acc $\sim$63–80\% & 3×15 m walk \\

%\cite{schlachetzki2017wearable} & Shoe-mounted IMUs & 190 PD + 101 HC & No ML split & PD vs HC, UPDRS & Signal processing pipeline (no ML) & Objective gait quantification & NR & 4×10 m walk + 180° turns \\

%\cite{benyoucef2025wearable} & Wearable sensors + video & 8 PD + 1 HC & Segment-level split & Windows labeled by subject & CNN-LSTM + RF & Binary classification & 100\% Acc, AUC=1.0 & 3–10 min total (5-s segments) \\

%\cite{haussler2025refinement} & 3 IMUs (feet + hip) & 19 total (14 analyzed) & No ML split & Video reviewed by clinician & Refined pFOG algorithm & FoG detection/prediction & Acc=71–89\% & NR \\

\end{longtable}

\noindent\footnotesize\textit{Abbreviations:} PD = Parkinson's Disease; HC = Healthy Controls; TUG = Timed Up and Go; AUC = Area Under ROC Curve.
\end{landscape}

\twocolumn